\PassOptionsToPackage{unicode}{hyperref}
\PassOptionsToPackage{hyphens}{url}
\PassOptionsToPackage{dvipsnames,svgnames,x11names}{xcolor}
\documentclass[
]{article}
\usepackage{amsmath,amssymb}
\usepackage{lmodern}
\usepackage{iftex}
\ifPDFTeX
  \usepackage[T1]{fontenc}
  \usepackage[utf8]{inputenc}
  \usepackage{textcomp} 
\else 
  \usepackage{unicode-math}
  \defaultfontfeatures{Scale=MatchLowercase}
  \defaultfontfeatures[\rmfamily]{Ligatures=TeX,Scale=1}
\fi
\IfFileExists{upquote.sty}{\usepackage{upquote}}{}
\IfFileExists{microtype.sty}{
  \usepackage[]{microtype}
  \UseMicrotypeSet[protrusion]{basicmath} 
}{}
\makeatletter
\@ifundefined{KOMAClassName}{
  \IfFileExists{parskip.sty}{%
    \usepackage{parskip}
  }{
    \setlength{\parindent}{0pt}
    \setlength{\parskip}{6pt plus 2pt minus 1pt}}
}{
  \KOMAoptions{parskip=half}}
\makeatother
\usepackage{xcolor}
\NewDocumentCommand\citeproctext{}{}
\NewDocumentCommand\citeproc{mm}{%
  \begingroup\def\citeproctext{#2}\cite{#1}\endgroup}
\makeatletter
 \let\@cite@ofmt\@firstofone
 \def\@biblabel#1{}
 \def\@cite#1#2{{#1\if@tempswa , #2\fi}}
\makeatother
\newlength{\cslhangindent}
\newlength{\csllabelwidth}
\newenvironment{CSLReferences}[2] 
 {\begin{list}{}{%
  \setlength{\itemindent}{0pt}
  \setlength{\leftmargin}{0pt}
  \setlength{\parsep}{0pt}
  \ifodd #1
   \setlength{\leftmargin}{\cslhangindent}
   \setlength{\itemindent}{-1\cslhangindent}
  \fi
  \setlength{\itemsep}{#2\baselineskip}}}
 {\end{list}}
\usepackage{calc}

\ifLuaTeX
\usepackage[bidi=basic]{babel}
\else
\usepackage[bidi=default]{babel}
\fi
\babelprovide[main,import]{american}

\def\languageshorthands#1{}
\ifLuaTeX
  \usepackage{selnolig}  
\fi
\IfFileExists{bookmark.sty}{\usepackage{bookmark}}{\usepackage{hyperref}}
\IfFileExists{xurl.sty}{\usepackage{xurl}}{} 
\hypersetup{
  pdftitle={GALFITools: A Python library to enhance GALFIT usage in
galaxy image modeling},
  pdfauthor={Christopher Añorve},
  pdflang={en-US},
  colorlinks=true,
  linkcolor={Maroon},
  filecolor={Maroon},
  citecolor={Blue},
  urlcolor={Blue},
  pdfcreator={LaTeX via pandoc}}

\title{GALFITools: A Python library to enhance GALFIT usage in galaxy
image modeling}

\definecolor{c53baa1}{RGB}{83,186,161}
\definecolor{c202826}{RGB}{32,40,38}

\usepackage[affil-it]{authblk}
\usepackage{orcidlink}
\author[1%
  \ensuremath\mathparagraph]{Christopher Añorve%
    \,\orcidlink{0000-0002-3721-8869}\,%
    }

\affil[1]{Facultad de Ciencias de la Tierra y el Espacio, Universidad
Autónoma de Sinaloa, México%
  }
\affil[$\mathparagraph$]{Corresponding author: %
}
\date{1 May 2025}

\begin{document}
\maketitle

\section{Summary}\label{summary}

Understanding how galaxies form and evolve requires measuring their
light distributions in images taken by telescopes. This process often
involves fitting mathematical models to galaxy images to extract
properties such as size, brightness, components, and shape. GALFIT is a
widely used tool for this purpose, but it requires careful preparation
of input files and interpretation of results, which can be a barrier to
efficient use.

GALFITools is a Python library that streamlines this workflow by
automating many of the tasks surrounding the use of GALFIT. These
include generating image masks, estimating sky background, modeling the
telescope's point spread function (PSF), and extracting physical
parameters from GALFIT outputs. The software is designed for researchers
and students who work with galaxy image modeling and aims to make the
process more reproducible, accessible, and scalable.

\section{Statement of need}\label{statement-of-need}

The analysis of galaxy morphology through image fitting is a fundamental
task in extragalactic astronomy. GALFIT (\citeproc{ref-peng02}{Peng et
al., 2002}, \citeproc{ref-peng10}{2010}) is a well-established tool that
performs parametric two-dimensional modeling of galaxy surface
brightness profiles. However, GALFIT does not provide utilities for
related tasks such as generating mask images, selecting stars for PSF
modeling, estimating initial fit parameters, or interpreting its output
formats in a structured way. These tasks are typically handled manually
or with ad-hoc scripts, which reduces reproducibility, becomes
time-consuming and error-prone, raises the barrier for new users, and
makes the application to large surveys difficult.

GALFITools (\citeproc{ref-anorve24}{Añorve, 2024}) addresses this gap by
offering a cohesive suite of Python-based tools that extend the
functionality of GALFIT. It facilitates the full modeling pipeline, from
input preparation to result interpretation. This includes routines to
construct PSFs, model galaxies via Multi-Gaussian Expansion (MGE,
\citeproc{ref-cappellari02}{Cappellari, 2002}), estimate sky
backgrounds, simulate galaxies, estimate initial parameters, generate
diagnostic plots, visualise models, and compute photometric parameters
from multiple Sérsic components such as effective radius, bar size and
Sérsic index (among others). These functionalities are available as
command-line tools, in line with GALFIT's command-line interface, but
can also be imported as Python modules.

GALFITools lowers the technical barriers for new users and increases
efficiency for experienced researchers. The package is designed for
astronomers who analyze galaxy images and require flexible, scriptable
tools that complement GALFIT analysis. While other packages such as
\texttt{Imfit} (\citeproc{ref-erwin15}{Erwin, 2015}) and \texttt{ProFit}
(\citeproc{ref-robotham16}{Robotham et al., 2016}) provide similar
modeling capabilities, GALFIT remains one of the most widely used tools
in the astronomical community for galaxy image modeling, supported by an
extensive user base and numerous legacy workflows. GALFITools is not a
replacement for GALFIT, but a complementary toolset specifically
tailored to its ecosystem. By reducing the technical overhead and
providing automation, GALFITools supports researchers in conducting
large-scale studies of galaxy structure and photometry with improved
efficiency and reproducibility.

\section{Acknowledgements}\label{acknowledgements}

The author acknowledges support from Universidad Autónoma de Sinaloa
through project PROFAPI 2022, with the project key A1009. GALFITools was
developed using PyScaffold. GALFIT itself is maintained by Chien Peng.

\section*{References}\label{references}
\addcontentsline{toc}{section}{References}

\phantomsection\label{refs}


\begin{CSLReferences}{1}{0}
\bibitem[\citeproctext]{ref-anorve24}
Añorve, C. (2024). \emph{{GALFITools v1.15.0}}.
\url{https://doi.org/10.5281/zenodo.13994492}

\bibitem[\citeproctext]{ref-cappellari02}
Cappellari, M. (2002). Efficient multi-gaussian expansion of galaxies.
\emph{Monthly Notices of the Royal Astronomical Society}, \emph{333}(2),
400--410. \url{https://doi.org/10.1046/j.1365-8711.2002.05412.x}

\bibitem[\citeproctext]{ref-erwin15}
Erwin, P. (2015). Imfit: A fast, flexible new program for astronomical
image fitting. \emph{The Astrophysical Journal}, \emph{799}(2), 226.
\url{https://doi.org/10.1088/0004-637X/799/2/226}

\bibitem[\citeproctext]{ref-peng02}
Peng, C. Y., Ho, L. C., Impey, C. D., \& Rix, H.-W. (2002). Detailed
structural decomposition of galaxy images. \emph{The Astronomical
Journal}, \emph{124}(1), 266--293. \url{https://doi.org/10.1086/340952}

\bibitem[\citeproctext]{ref-peng10}
Peng, C. Y., Ho, L. C., Impey, C. D., \& Rix, H.-W. (2010). Detailed
decomposition of galaxy images. II. Beyond axisymmetric models.
\emph{The Astronomical Journal}, \emph{139}(6), 2097--2129.
\url{https://doi.org/10.1088/0004-6256/139/6/2097}

\bibitem[\citeproctext]{ref-robotham16}
Robotham, A. S. G., Taranu, D. S., Tobar, R., Moffett, A., \& Driver, S.
P. (2016). ProFit: Bayesian profile fitting of galaxy images.
\emph{Monthly Notices of the Royal Astronomical Society}, \emph{466}(2),
1513--1541. \url{https://doi.org/10.1093/mnras/stw3039}

\end{CSLReferences}
\end{document}